# The prediction of the quality of results in Logic Synthesis using Transformer and Graph Neural Networks

Chenghao Yang*

Ningbo University, China, 1911082192@nbu.edu.cn.

Zhongda Wang

University of New South Wales, Australia, z5234782@ad.unsw.edu.au.

Yinshui Xia

Ningbo University, China, xiayinshui@nbu.edu.cn

Zhufei Chu

Ningbo University, China, chuzhufei@nbu.edu.cn

In the logic synthesis stage, structure transformations in the synthesis tool need to be combined into optimization sequences and act on the circuit to meet the specified circuit area and delay. However, logic synthesis optimization sequences are time-consuming to run, and predicting the quality of the results (QoR) against the synthesis optimization sequence for a circuit can help engineers find a better optimization sequence faster. In this work, we propose a deep learning method to predict the QoR of unseen circuit-optimization sequences pairs. Specifically, the structure transformations are translated into vectors by embedding methods and advanced natural language processing (NLP) technology (Transformer) is used to extract the features of the optimization sequences. In addition, to enable the prediction process of the model to be generalized from circuit to circuit, the graph representation of the circuit is represented as an adjacency matrix and a feature matrix. Graph neural networks(GNN) are used to extract the structural features of the circuits. For this problem, the Transformer and three typical GNNs are used. Furthermore, the Transformer and GNNs are adopted as a joint learning policy for the QoR prediction of the unseen circuit-optimization sequences. The methods resulting from the combination of Transformer and GNNs are benchmarked. The experimental results show that the joint learning of Transformer and GraphSage gives the best results. The Mean Absolute Error (MAE) of the predicted result is 0.412.

---

* C. Yang, Y. Xia and Z. Chu are with the Faculty of Electrical Engineering and Computer Science, Ningbo University, Ningbo 315211, P.R. Z. Wang is with the Faculty of Computer Science and Engineering, University of New South Wales, G30/ 9 Rosebery Ave. Zetland 2017 NSW, AUS.

## 1 INTRODUCTION

As the intermediate layer of most electronic design automation (EDA) processes, logic synthesis is defined as the process of optimizing a given Boolean network and mapping it to a gate-level netlist while optimizing quality-of-results (QoR). In the pre-mapping phase of logic synthesis, researchers have developed several structural transformations to improve the QoR of the circuit [1]. Circuits are represented as And-Inverter Graphs (AIG) in the state-of-the-art academic logic synthesis tool ABC [2]. These technology-independent structural transformations are used to reduce the size of the graph to achieve circuit optimization. As the complexity of the design and technology grows, developers need to combine these structural transformations into logic synthesis optimization sequences (hereafter referred to as optimization sequences) based on their knowledge and after extensive iterative testing to meet the optimization goals. However, logic synthesis techniques lack predictability, and hence it is difficult for developers to know the performance of optimization sequences. Hence, predicting the performance of optimization sequences without actual testing is of great value [3].

Deep learning (DL) has developed rapidly in recent years. Through the powerful fitting ability of deep neural networks, it can learn a large number of mapping relationships between inputs and outputs, without requiring a precise mathematical representation between inputs and outputs. Hence, logic synthesis researchers aim to take advantage of advances in DL to facilitate the convergence rate of the logic synthesis phase.

In [4], [5], the logic synthesis optimization process is modeled as a Markovian decision process, and reinforcement learning algorithms incorporating graph neural networks are used to explore the design space. In [6], circuit area under delay constraints is minimized using reinforcement learning algorithm A2C. In [7], a milestone step by releasing the dataset is taken. Yu et al. [8] mapped Logic Synthesis Optimization to a multi-classification problem. The main idea is to use convolutional neural networks (CNN) to classify the logic synthesis optimization sequences encoded as one-hot to the best and worst. In [9], QoR prediction of optimization sequences for a specific design is implemented using Long Short-Term Memory (LSTM) networks. However, in [8], [9], the input to the neural network contains information about the optimization sequence only, and the neural network cannot learn information about the structure of the circuit. Hence, the trained network can only work for a specific circuit. In this work, a framework for predicting the QoR of unseen **circuit-optimization sequence pairs** is proposed. Specifically, graph neural networks (GNNs) are used to process adjacency and feature matrices containing circuit structure information to learn circuit structure features. The GNNs utilized are Graph Convolutional Network (GCN) [13], Graph Attention Network (GAT) [14], and GraphSage [15]. In addition, optimization sequences are treated as sentences in natural language and modeled as learnable embedding vectors. Advanced natural language processing (NLP) techniques are used to learn the features of the optimization sequences.

The main contributions of this paper are as follows:

- An optimization sequence feature extractor based on the NLP algorithm Transformer is proposed. The input is an optimization sequence that has been transformed into a vector by the embedding method, while the output is a feature vector extracted from the optimization sequence that can be used for downstream QoR prediction tasks.
- A circuit feature extractor is developed based on three typical graph neural networks, functioning to extract features of the circuit structure where the circuits are represented as adjacency matrix and feature matrix.
- A joint learning policy is proposed for the optimization sequence feature extractor and circuit feature extractor, enabling the model to predict the performance of unseen circuit-optimization sequence pairs. Furthermore, experiments are conducted on the nine methods resulting from the combination, and suggestions for their uses are given for different scenarios, providing a basis for experimental analysis of subsequent related problems in logic synthesis.

- The code for this work will be open source at https://github.com/Chenghao-Yang/QoR-Prediction.

## 2 FUNDAMENTALS

To understand the potential of deep learning for the problem of predicting the QoR of optimization sequences, preliminary knowledge about Transformer [12] and graph neural networks is presented. In addition, two classical NLP feature extractors CNN [10] and LSTM [11], which are used as baselines, are also introduced.

### 2.1 CNN Basics

A convolutional neural network contains a feature extractor consisting of a convolutional layer and a sub-sampling layer. A convolutional layer typically contains several feature maps, and each feature map consists of several rectangularly arranged neurons. A neuron is connected to only part of the neighboring neurons. The neurons in the same feature map share weights, where the shared weights are the convolutional kernel. The convolutional kernel is generally initialized in the form of a random algebraic matrix, and the kernel will learn to obtain reasonable weights during the training process of the network. The direct benefit of shared weights (convolution kernels) is to reduce the parameters in each layer of the network, while reducing the risk of overfitting. Subsampling, also called pooling, is generally available in the form of mean pooling and max pooling. Subsampling can be seen as a special kind of convolution process. Convolution and subsampling greatly simplify the model complexity and reduce the parameters of the model. The process of extracting optimization sequence features using CNN is shown in Section 4.1.3.

### 2.2 LSTM Basics

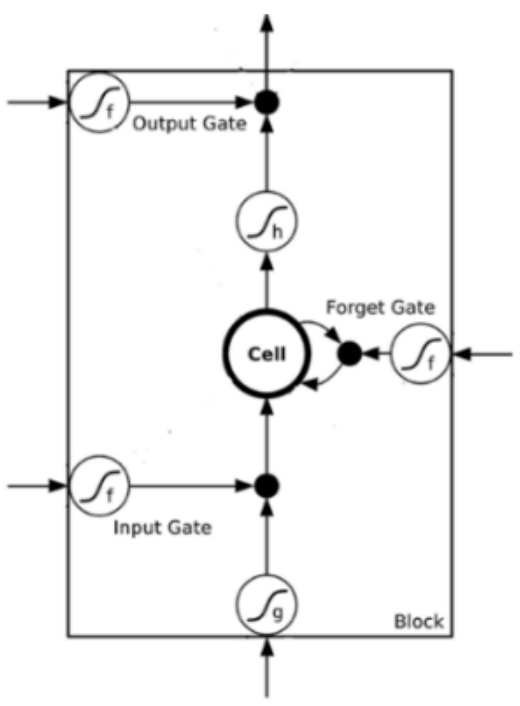

Figure 1: LSTM unit illustration.

Long short-term memory (LSTM) [11] is a recurrent neural network (RNN) architecture. It is to solve long-term dependency problems by adding control gates to recurrent units. As shown in Figure 1, a common architecture of an LSTM unit consists of cell, input gate, output gate, and forget gate. Input gates receive inputs from sequences and other units of the neural network and are trained to add information to the cell at an appropriate time. Similar to the input gate, the output gate learns to output the information from the cell at the appropriate moment. The forget gate learns the appropriate moment to remove information from the cell when it is no longer useful.

These gates consist of a sigmoid layer and a pointwise multiplication to allow selective passage of information.

## 2.3 Transformer Basics

The Transformer network architecture was proposed by Ashish Vaswani et al [12] and used for machine translation tasks. In contrast to previous network architectures, the encoder and decoder do not use a network architecture such as RNN or CNN, but an architecture that relies entirely on attention mechanisms. As the model processes each input vector, self-attention helps the model to look at other positions in the input sequence to find relevant cues for better encoding. The basic architecture of the Transformer is shown in Figure 2, where the left half is the Encoder part and the right half is Decoder part. The transformer has six layers of such structure. In each layer, the Encoder converts the input sequence into an encoded representation, while the Decoder produces a new sequence as the output based on the encoded representation.

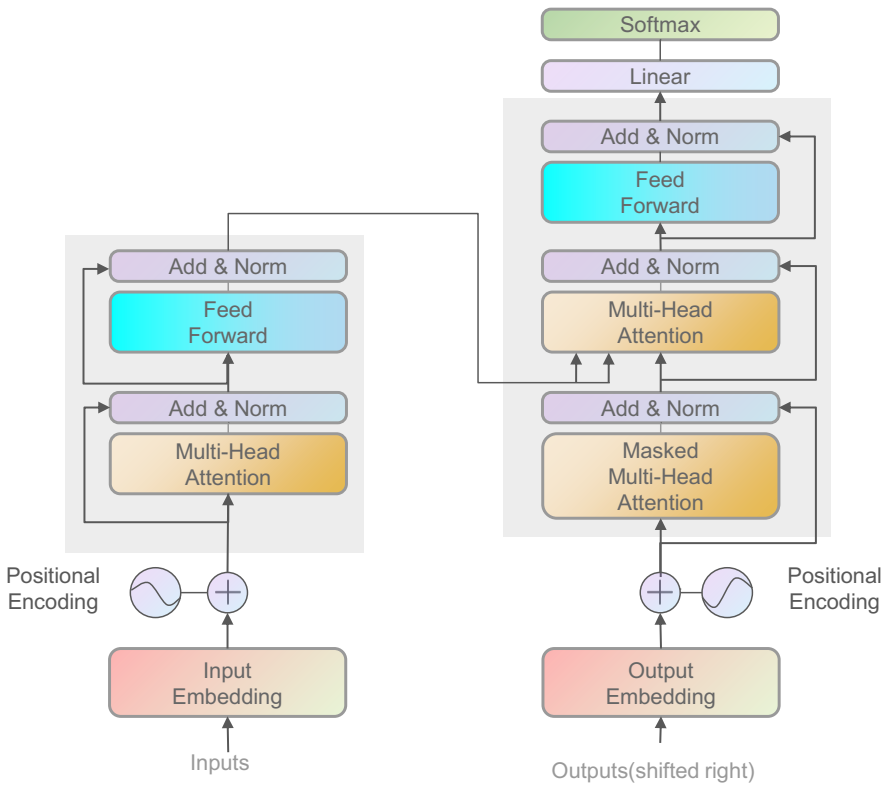

Figure 2: Transformer Architecture.

Since this work does not require converting the input optimization sequence to another sequence, the direct application of the transformer is not applicable. How to extract features of the optimization sequence using partial components of the transformer is shown in 4.1.2.

## 2.4 Graph Neural Networks

GNNs are neural networks that operate directly on graph data structures. Take learning the embedding vector of each node in the input graph for example. The features of node $v$ are represented by $\mathbf{x}_v$, and the network learns to represent the node by a d-dimensional vector (state) $\mathbf{h}_v$, which contains information about its neighborhood.

$$\mathbf{h}_v = f\left(\mathbf{x}_v, \mathbf{x}_{co[v]}, \mathbf{h}_{ne[v]}, \mathbf{x}_{ne[v]}\right) \tag{1}$$

where $\mathbf{x}_{co[v]}$ denotes the features of the edge connected to $v$, $\mathbf{h}_{ne[v]}$ the embedding features of the neighboring nodes of $v$, $\mathbf{x}_{ne[v]}$ the features of the neighboring nodes of $v$. $f$ is the transfer function that projects these inputs into the d-dimensional space. Since a unique solution for $\mathbf{h}_v$ is being solved, Banach Fixed Point Theorem can be applied and the above equation can be rewritten as an iterative update procedure.

$$\mathbf{H}^{t+1} = F(\mathbf{H}^t, \mathbf{X}) \tag{2}$$

$\mathbf{H}$ and $\mathbf{X}$ denote all connections of $\mathbf{h}$ and $\mathbf{x}$, respectively. The output of GNN is calculated by passing state $\mathbf{h}_v$ as well as feature $\mathbf{x}_v$ to the output function $g$.

$$\mathbf{o}_v = g(\mathbf{h}_v, \mathbf{x}_v) \tag{3}$$

Both $f$ and, here, $g$ can be interpreted as feedforward fully connected (FC) neural networks.

GNN basic tasks are classified according to graph structure elements and can be divided into nodes, edges, subgraphs, and graph-related tasks. In this paper, GNNs are used to learn graph embeddings of circuits, which contain structural features of the circuits and can be used for downstream QoR prediction tasks.

## 3 MOTIVATION

There is a correlation between the structural transformations used to optimize the circuit. In other words, the use of structural transformations in conjunction with each other can produce different optimization effects. For example, in ABC, zero-cost replacements (rewrite -z) do not immediately reduce the number of nodes in the AIG graph, and other structural transformations are required to complete the optimization. Self-attention in Transformer processes each vector in the input sequence by computing the correlation of that vector with other vectors in the sequence and looking for relevant clues. Hence, the architecture of the transformer is well suited for extracting the features of optimization sequences. In addition, in ABC [2], circuits are represented as graphs (e.g. AIG), and a significant number of structure transformation algorithms are based on graph structures. In previous work [8], [9], the input to the neural network only contains information about the optimization sequence, and it is not possible to establish the correlation between the optimization sequence and the circuit structure. Hence, this paper combines a graph neural network with a joint policy to learn both optimization sequence features and circuit structure features. We aim to achieve QoR predictions for unseen Circuit-Optimization sequences.

## 4 APPROACH

The input and the architecture of the model for neural networks are mainly described. The input is the optimization sequence, which is converted into a vector representation after the embedding layer. The neural network architecture consists of an optimization sequence feature extractor and a circuit feature extractor which are used to predict the QoR of the optimization sequence for different circuits after adopting a joint learning policy.

### 4.1 Optimization Sequence Feature Extractor

The implementation of an optimization sequence feature extractor based on the Transformer is presented here. In addition, for a comprehensive comparison, the classical LSTM and the CNN used in [8], [9] are implemented as well.

#### *4.1.1 Optimization Sequence Embedding*

The input optimization sequence in this paper consists of seven ABC structure transformations, specifically **refactor, refactor -z, rewrite, rewrite -z, resub, resub -z, balance,** and a sequence of these seven structure transformations is assumed to be **[rewrite, resub, refactor, balance]**. In the following, we will show how to convert the optimization sequence into a vector that can be processed by the neural network.

First, the original seven structural transformations are used as dictionaries. Each structural transformation is considered as a word in natural language and is assigned an index of 0 to 6, respectively. To make the neural network learn better, the embedding matrix is further created. Each index is assigned three latent factors, that is, represented by a vector of length 3. For better illustration, we randomly initialize an embedding matrix for seven structural transformations:

$$
\begin{bmatrix}
-0.4093 & -1.1011 & 0.0790 \\
-0.2704 & 0.0708 & 0.6557 \\
-0.5706 & -0.2703 & 2.2453 \\
0.6731 & -0.6557 & -0.9846 \\
-1.1936 & -0.0705 & 0.3704 \\
0.2741 & 0.4531 & 2.5046 \\
0.4327 & -0.9486 & 2.2643
\end{bmatrix}
$$

The row numbers of the matrix correspond to the index values, and each row vector represents the embedding vector corresponding to the index. As an example, an optimization sequence **[rewrite, resub, refactor, balance]** can be considered as a sentence composed of a dictionary. By querying the indexes in the dictionary, the optimization sequence can be transformed to [2, 4, 0, 6]. By querying rows 2, 4, 0, 6 of the embedding matrix by indexes, respectively, the optimization sequence is represented as the following embedding vector:

$$
\begin{bmatrix}
-0.5706 & -0.2703 & 2.2453 \\
-1.1936 & -0.0705 & 0.3704 \\
-0.4093 & -1.1011 & 0.0790 \\
0.4327 & -0.9486 & 2.2643
\end{bmatrix}
$$

Each row represents a structural transformation performed at a one-time step. Note that not every structural transformation is replaced by a vector, but is replaced by the index used to find the vectors in the embedding matrix. The values of the entire embedding matrix are used as part of the training, as the model is trained, the embedding vector gets the appropriate values.

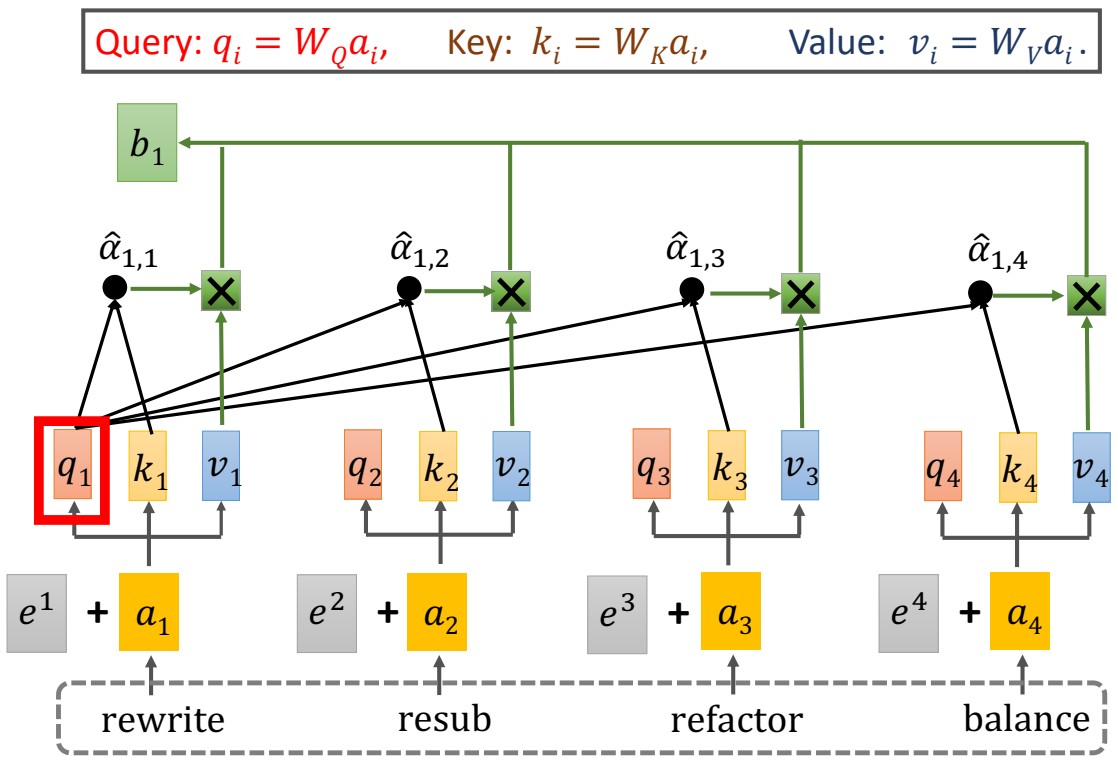

Figure 3: optimization sequence feature extractor based on self-attention.

*4.1.2 Transformer-based optimization sequence feature extractor*

Here, how to use the partial modules in the transformer for optimization sequence feature extraction will be presented.

**a) Computation of self-attention**

As the core part of the transformer, optimization sequence feature extraction using the self-attention mechanism is introduced first. As shown in Figure 3, the structural transformations in the optimization sequence [rewrite, resub, refactor, balance] are transformed into vectors $a_1$, $a_2$, $a_3$, and $a_4$ by the optimization sequence embedding in 4.1.1 and added with the position information $e^i$, respectively. For position information $e^i$, the method proposed in [16] is used in this paper, a simpler method than the original transformer. This approach is similar to that of generating word vectors, where the position encoding is first initialized and then put into the training process to train a position vector for each position.

The next step is to compute self-attention. In the first step, each input vector $a_i$ is multiplied with the parameter matrices $W_Q$, $W_K$, and $W_V$ to obtain the query vector $q_i$, the key vector $k_i$, and the value vector $v_i$, respectively. The second step is to compute the score. Figure 3 demonstrates the process of computing the self-attention vector for the first structural transformation 'rewrite'. During the computation, each structural transformation in the input optimization sequence needs to be used to score 'rewrite'. This score determines how much attention should be given to the rest of the input sequence when encoding the structural transformation 'rewrite'. It is computed by the dot product of the query vector of 'rewrite' with the key vector of each structural transformation. For example, the first score is the dot product of $q_1$ and $k_1$, and the second score is the dot product of $q_1$ and $k_2$. To make the gradient smoother, the obtained score is divided by $d$ as follows:

$$\alpha_{1,i} = q_1 \cdot \frac{k_i}{\sqrt{d}} \tag{4}$$

where $d$ is the dimensionality of $q$ and $k$. Then, the softmax score is obtained by normalizing the scores using the softmax function with the following equation:

$$\hat{\alpha}_{1,i} = \frac{exp(\alpha_{1,i})}{\sum_j exp(\alpha_{1,j})} \tag{5}$$

This makes the scores all positive and sum to 1. In the third step, each value vector $v_i$ is multiplied by the softmax score and summed, which gives the output of the self-attention layer at the first position in the sequence, $b_1$. A similar calculation for the remaining $a_2$, $a_3$, and $a_4$ gives $b_2$, $b_3$, and $b_4$. The final self-attention matrix is formed by combining all the $b_i$. In practice, the computation is performed as a matrix that can be accelerated by the GPU. To extend the ability of the model to focus on different positions, the multi-headed attention mechanism is further utilized.

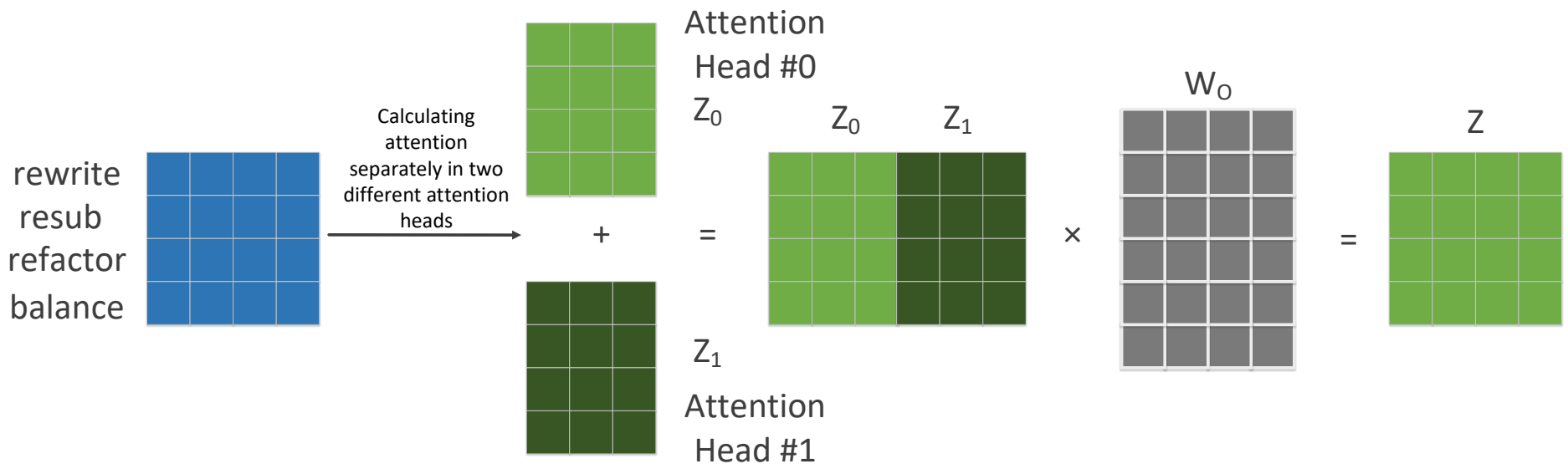

Figure 4: The process of computing multi-headed self-attention.

In this paper, two heads are used and each head has an independent query, key, and value weight matrix. As shown in Figure 4, each attention head independently performs the same operation as the single head in the previous paragraph to obtain the corresponding $b_i$. Matrices $Z_0$ and $Z_1$ are formed by combining the corresponding $b_i$. After that, $Z_0$ and $Z_1$ are concatenated and multiplied by an additional weight matrix $W_O$ to obtain the final self-attention matrix Z.

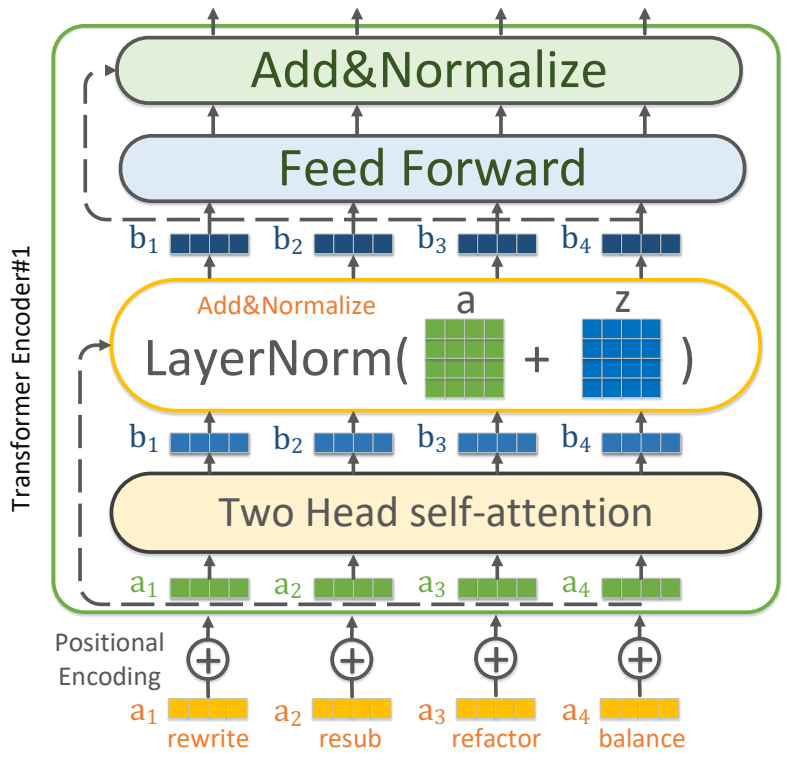

Figure 5: Transformer-based optimization sequence feature extractor.

**b) Transformer-based extractor**

The transformer-based optimization sequence feature extractor is completely demonstrated here. A block of the optimization sequence feature extractor is shown in Figure 5. To extract sequence features, the input optimization sequence is first transformed into vectors $a_i$ by the optimization sequence embedding in 4.1.1 These vectors are added with the corresponding position encoding to become the new vector $a_i$. After that, the vector $a_i$ is passed into the two head self-attention in section 4.1.2.a for computing the self-attention matrix Z. The next layer is the residual connection and layer normalization module. The residual connection is adding the self-attention matrix Z with the corresponding input vector $a_i$, to prevent network degradation. The added matrix is layer normalized [17] to normalize the activation values for that layer. The output after layer normalization is linearly transformed by feed-forward layer. Finally, the input and output of the feedforward network are residual-connected and then layer-normalized to obtain the final output. In this paper, the optimization sequence feature extractor consists of three blocks stacked together. In addition, the output of the extractor does dimensional scaling through a linear layer.

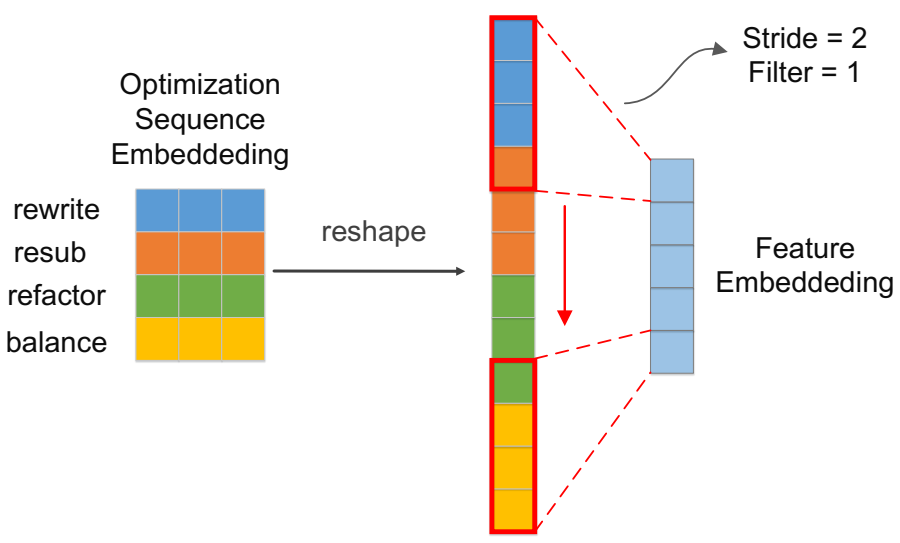

Figure 6: Extraction of optimization sequence features using CNN.

*4.1.3 The optimization sequence feature extractor based on CNN and LSTM*

As a comparison, the CNN and LSTM, which were used in [8], [9] respectively, are implemented in this paper. An example of a CNN-based optimization sequence feature extractor is shown in Figure 6. The embedding vector representation of the optimization sequence is reshaped and features are extracted using a convolution kernel. Specifically, the convolution kernel slides top-down in steps of two. With each slide, the convolution kernel is dot producted with the embedded vectors within the window of the convolution kernel, and the final sequence features consist of the results of each dotted product.

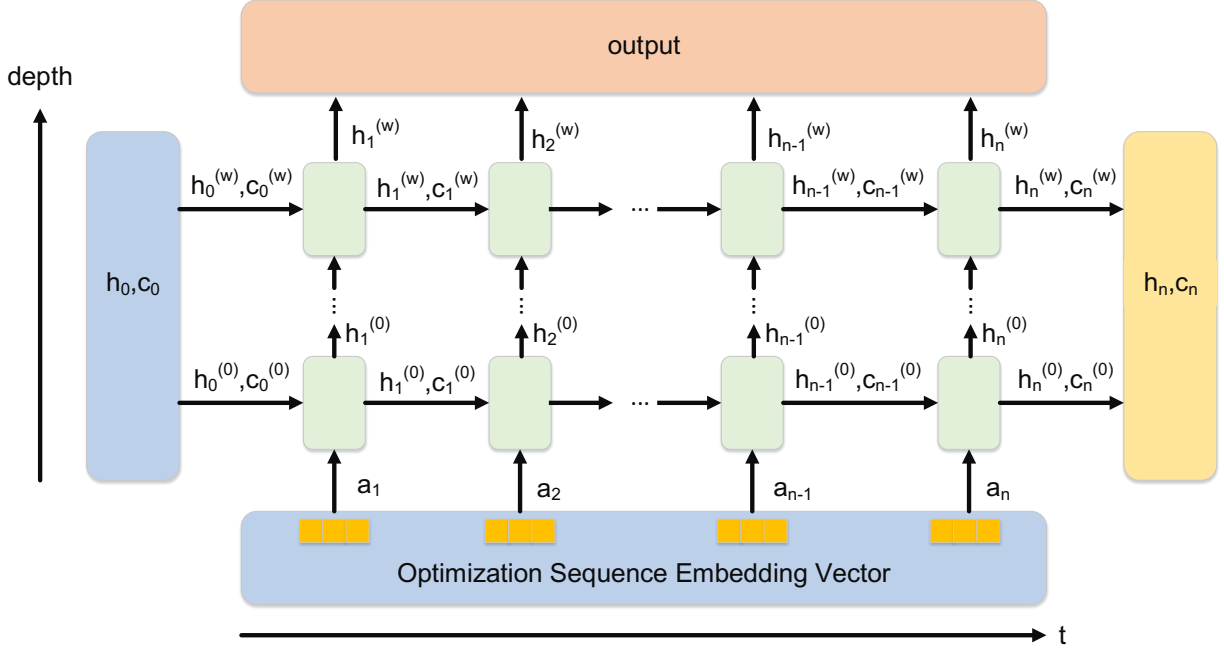

Figure 7: Feature extraction for optimization sequences using LSTM.

Detailed parameters are shown in Table 2. The example procedure using LSTM to extract features of the optimization sequence is shown in Figure 7. The optimization sequence is considered as time-series data input to the LSTM network, and each unit in the LSTM updates the cell state and output hidden state based on the existing input and the output of the previous unit. In this paper, the last hidden state of each layer is used as the embedding feature of the whole optimization sequence. Detailed parameters are shown in Table 2.

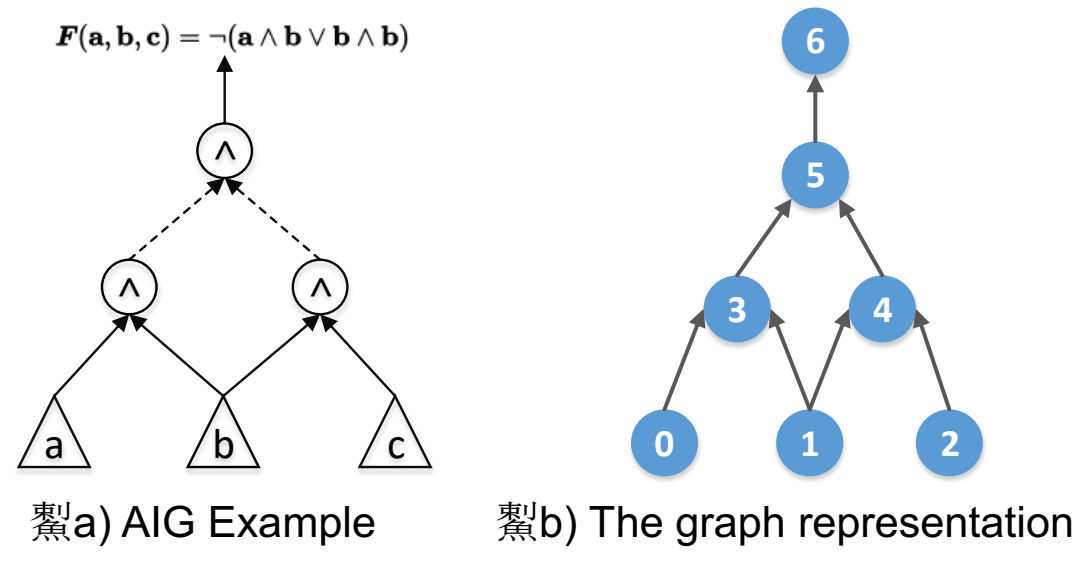

Figure 8: AIG example of the logic $\boldsymbol{F}(\mathbf{a},\mathbf{b},\mathbf{c}) = \neg(\mathbf{a} \wedge \mathbf{b} \vee \mathbf{b} \wedge \mathbf{b})$.

### 4.2 Circuit feature extractor

Since the circuit can be represented as a directed acyclic graph, the graph neural network can be used to extract information about the structural features of the circuit. Figure 8 shows an example of AIG (8-a), which has three inputs and one output. The solid and dashed lines represent the buffer and inverter, respectively, and the node in the middle is the AND logic. In Figure 8-b, the AIG is converted to a graph representation. To distinguish the attributes of different nodes, different feature vectors are added to the nodes in the graph representation. For illustration, the first iteration of the simplified propagation rule will be shown below.

$$A = \begin{bmatrix} 0 & 0 & 0 & 1 & 0 & 0 & 0 \\ 0 & 0 & 0 & 1 & 1 & 0 & 0 \\ 0 & 0 & 0 & 0 & 1 & 0 & 0 \\ 0 & 0 & 0 & 0 & 0 & 1 & 0 \\ 0 & 0 & 0 & 0 & 0 & 1 & 0 \\ 0 & 0 & 0 & 0 & 0 & 0 & 1 \\ 0 & 0 & 0 & 0 & 0 & 0 & 0 \end{bmatrix}, F = \begin{bmatrix} 0 & 0 \\ 0 & 0 \\ 0 & 0 \\ 2 & 0 \\ 2 & 0 \\ 2 & 2 \\ 1 & 0 \end{bmatrix} \rightarrow X = A \times F = \begin{bmatrix} 2 & 0 \\ 4 & 0 \\ 2 & 0 \\ 2 & 2 \\ 2 & 2 \\ 1 & 0 \\ 0 & 0 \end{bmatrix}$$

where A is the adjacency matrix representing the graph connectivity, $F$ is the feature matrix representing the node attributes, and the feature addition rules follow [7]. Each node is initialized with a feature vector of length 2. The first dimension of the feature matrix represents the node type, where the input, output and AND gates are denoted as 0, 1 and 2, respectively. The second dimension is the number of inverters of the node inputs. The matrix $X$ obtained after multiplying A and $F$ represents the node embedding after propagation, e.g., the row vector [2,0] in row 0 of $X$ stands for node 0 after propagation. The following will describe how three typical graph neural networks learn circuit features.

#### *4.2.1 GCN-based circuit feature extractor*

A graph convolutional neural network, much like a regular convolution, creates convolutions by aggregating nodes and their neighbors to form new nodes. Like any other neural network, GCNs can be stacked with multiple layers. For each layer, the aggregation of nodes, also known as convolution, can be expressed as an equation:

$$H^{(l+1)} = \sigma\left(\tilde{D}^{-\frac{1}{2}}\tilde{A}\tilde{D}^{-\frac{1}{2}}H^{(l)}W^{(l)}\right) \tag{6}$$

$\tilde{A}$ is the adjacency matrix with the addition of a self-loop, which makes each node include its own features in the calculation. $\tilde{D}$ is the degree matrix of $\tilde{A}$, which is used to normalize nodes with larger degrees. $H^{(l)}$ denotes the node embedding for the output of layer $l$, which is also the input to the next layer. Initially, $H^{(0)} = F$, and $F$ is the feature matrix of the node. $W^{(l)}$represents the weight matrix of the $l$-th layer. $\sigma$ is the activation function, e.g., ReLU.

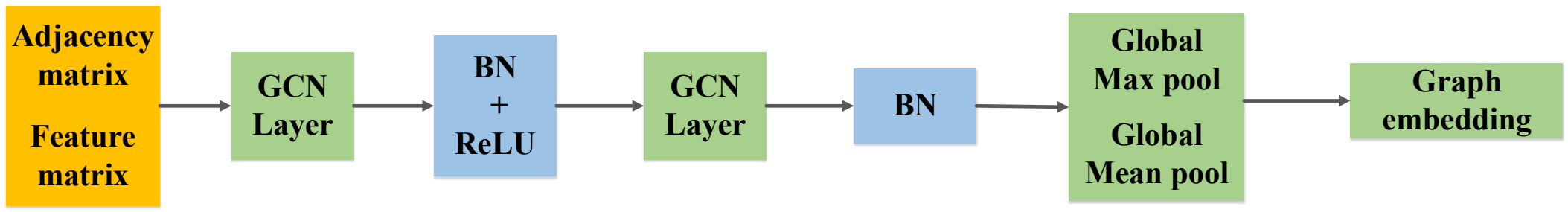

Figure 9: GCN-based circuit feature extractor.

The GCN-based circuit feature extractor used in this paper is shown in Figure 9. First is the input, where each vector representing the node type is converted to a 1 × 3 vector by the embedding layer, and then is concatenated with a vector representing the number of inverters for each node input to become a 1 × 4 node feature vector. Two GCN layers are also included, enabling each node to learn node embeddings from their neighbors' neighbor nodes. Each GCN layer is followed by batch normalization (BN) which helps to improve the training speed and accuracy. The activation function ReLU is used to increase the nonlinear properties of the network. Finally, the graph embedding consists of global max pooling and global mean pooling of node embeddings. The final graph embeddings will be used as extracted circuit features for downstream QoR prediction tasks.

#### *4.2.2 GAT-based circuit feature extractor*

The weights on the edges of the GCN are fixed at the time of aggregation for each convolution. GAT introduces the attention mechanism to let the model learn the weight assignment. The weight learning formula is as follows:

$$\alpha_{ij} = \frac{\exp\left(\mathrm{LeakyReLU}\left(\vec{\mathbf{a}}^T\left[\mathbf{W}\vec{h}_i \parallel \mathbf{W}\vec{h}_j\right]\right)\right)}{\sum_{k\in\mathcal{N}_i}\exp\left(\mathrm{LeakyReLU}\left(\vec{\mathbf{a}}^T\left[\mathbf{W}\vec{h}_i \parallel \mathbf{W}\vec{h}_k\right]\right)\right)} \tag{7}$$

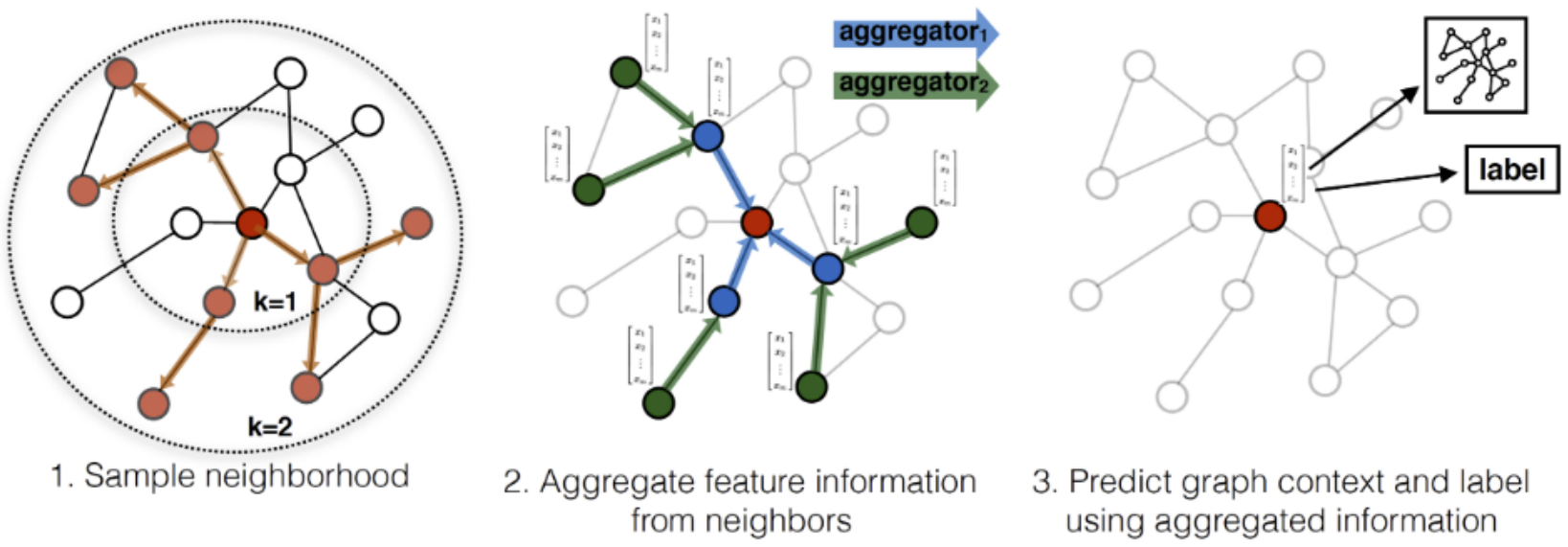

Figure 10: Graphsage schematic. Image cited from [15].

where $\alpha_{ij}$ denotes the attention coefficient between the $i$-th node and the $j$-th node, $\vec{h}_i$ is the node feature, and the other $\vec{\mathbf{a}}^T$, $\mathbf{W}$ are the learnable model parameters. The formulation first concatenates the representations of the target and source nodes and computes the correlation through the parameter network $\vec{\mathbf{a}}^T$. After that, the correlation is passed through the activation function leakyReLU. Finally, the computed results are normalized by the softmax function to obtain the attention score $\alpha_{ij}$. Feature aggregation is performed according to the attention score with the following equation:

$$\vec{h}_i' = \sigma\left(\sum_{j \in \mathcal{N}_i} \alpha_{ij} \mathbf{W} \vec{h}_j\right) \tag{8}$$

The aggregation process is equivalent to weighted summation. In this paper, the GAT-based circuit feature extractor is adapted from the module in Figure 9. Specifically, the GCN layer in it is replaced by the GAT layer, which means the aggregation process is changed to a weighted summation based on the attention coefficients. In addition, the activation function is changed from ReLU to elu.

#### *4.2.3 GraphSage-based circuit feature extractor*

In GraphSage, the feature aggregation process is represented as a function. The (structural and feature) information of nodes is passed from point to point. Through the aggregation function, a node can aggregate the information of its neighbors and update the information of the current node through the update function (neural network). In this paper, we use MEAN aggregate with the following equation:

$$h_v^k \leftarrow \sigma(\boldsymbol{W} \cdot \mathrm{MEAN}(\{h_v^{k-1}\} \cup \{h_u^{k-1}, \forall u \in N(v)\}) \tag{9}$$

The mean aggregator concatenates the $k-1$-th layer vectors of the target and neighbor nodes, then operates to find the mean value for each dimension of the vector, and does a nonlinear transformation of the obtained result to produce the node embedding of the $k$-th layer for the target node. In addition, as shown in Figure 10, to save computational resources, GraphSage allows sampling a certain number of neighboring nodes for each node as the nodes to be aggregated information. In the GraphSage-based circuit feature extractor in this paper, the GCN layer in Fig. 9 is replaced with GraphSage.

### 4.3 Joint learning policy

To predict the QoR of unseen circuit-optimization sequence pairs, optimization sequence feature extractor and circuit feature extractor are adopted as a joint policy. As shown in Figure 11, first, the optimization sequence and the AIG graph are passed into the sequence feature extractor and the circuit feature extractor, respectively.

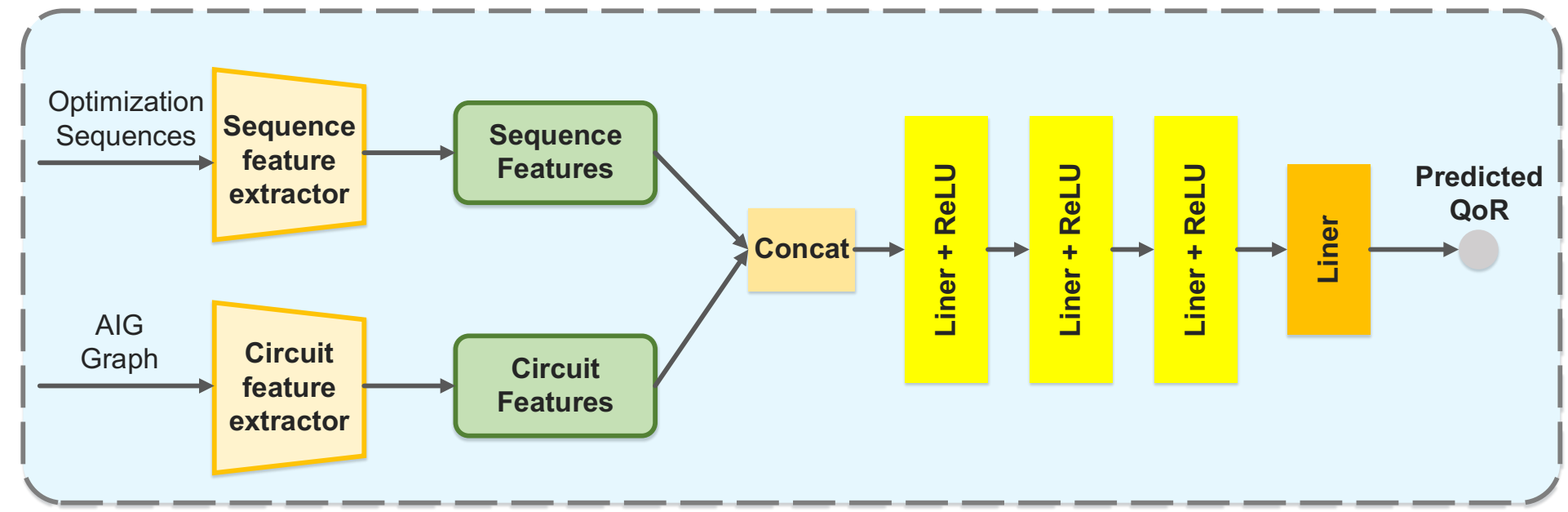

Figure 11: Sequence feature extractor and circuit feature extractor adopted a joint learning policy for predicting the QoR of optimization sequences.

Table 1: Hyperparameters of the circuit feature extractor. In the GAT-based module, both GAT layers use two-head attention. Among them, the two-head attention of Layer1 is concatenated and the two-head attention of Layer2 is averaged.

| GNN Type | Input | Layer1 | Layer2 | Pool | Output |
|---|---|---|---|---|---|
| GCN | 4 | 64 | 64 | Max+Mean | 128 |
| GAT | 4 | 32×2 | 64 | Max+Mean | 128 |
| GraphSage | 4 | 64 | 64 | Max+Mean | 128 |

Table 2: Hyperparameters of the optimization sequence feature extractor. The dimension of batch size is omitted.

| | **Transformer** | | | | | |
|---|---|---|---|---|---|---|
| **Sequence feature extractors** | Input | Num_Head | Dim_feedforward | Num_layers | Linear | Output |
| | (20, 4) | 2 | 32 | 3 | 50 | 50 |
| | **LSTM** | | | | | |
| | Input | Hidden_Size | Num_layers | Output | | |
| | (20, 3) | 64 | 2 | 64 | | |
| | **CNN** | | | | | |
| | Input | Filters | Kernels | Stride | Output | |
| | 60 | 4 | 21, 24, 27, 30 | 3 | 50 | |

Table 3: Hyperparameters of the fully connected layer. The input dimensions are divided into two types, 178: Transformer/CNN + GNN , 192: LSTM + GNN.

| **FC Stack** | Input | Linear1 | Linear2 | Linear3 | Linear4 | Dropout |
|---|---|---|---|---|---|---|
| | 178/ 192 | 512 | 256 | 256 | 1 | 0.2 |

Then, the sequence features and circuit features output from the two extractors are concatenated together. The concatenated features are passed through three linear layers with the activation function ReLU. Finally, a linear layer outputs the predicted QoR. With such a structure, the neural network can learn features of both the optimization sequence and the circuit and link them together. This allows predictions to be passed from circuit to circuit rather than being restricted to a specific circuit. The detailed parameters of the network are presented in Tables 1, 2, and 3.

Table 4: Characteristics of the circuits in the dataset OpenABC-D [7] used in this paper. Primary Inputs (PI), Primary outputs (PO), Nodes (N), Edges (E), Inverted edges (I), Netlist Depth (D).

| Circuit | Characteristics of Benchmarks | | | | | |
|---|---|---|---|---|---|---|
| | PI | PO | N | E | I | D |
| spi [21] | 254 | 238 | 4219 | 8676 | 5524 | 35 |
| i2c [21] | 177 | 128 | 1169 | 2466 | 1188 | 15 |
| ss_pcm [21] | 104 | 90 | 462 | 896 | 434 | 10 |
| usb_phy [21] | 132 | 90 | 487 | 1064 | 513 | 10 |
| sasc [21] | 135 | 125 | 613 | 1351 | 788 | 9 |
| wb_dma [21] | 828 | 702 | 4587 | 9876 | 4768 | 29 |
| simple_spi [21] | 164 | 132 | 930 | 1992 | 1084 | 12 |
| pci [21] | 3429 | 3157 | 19547 | 42251 | 25719 | 29 |
| wb_conmax [21] | 2122 | 2075 | 47840 | 97755 | 42138 | 24 |
| ac97_ctrl [21] | 2339 | 2137 | 11464 | 25065 | 14326 | 11 |
| mem_ctrl [21] | 1187 | 962 | 16307 | 37146 | 18092 | 36 |
| des3_area [21] | 303 | 64 | 4971 | 10006 | 4686 | 30 |
| aes [21] | 683 | 529 | 28925 | 58379 | 20494 | 27 |
| sha256 [22] | 1943 | 1042 | 15816 | 32674 | 18459 | 76 |
| aes_xcrypt [23] | 1975 | 1805 | 45840 | 93485 | 36180 | 43 |
| aes_secworks[24] | 3087 | 2604 | 40778 | 84160 | 45391 | 42 |
| fir [22] | 410 | 351 | 4558 | 9467 | 5696 | 47 |
| iir [22] | 494 | 441 | 6978 | 14397 | 8596 | 73 |
| tv80 [21] | 636 | 361 | 11328 | 23017 | 11653 | 54 |
| tiny_rocket [26] | 4561 | 4181 | 52315 | 108811 | 67410 | 80 |
| fpu [25] | 632 | 409 | 29623 | 59655 | 37142 | 819 |
| dynamic_node[26] | 2708 | 2575 | 18094 | 38763 | 23377 | 33 |

Table 5: Experimental results of the joint model of circuit feature extractor and optimization sequence feature extractor for the QoR prediction task (the lower the better). Results are averaged over 3 runs with 3 different seeds.

| Model | Transformer | | CNN | | LSTM | |
|---|---|---|---|---|---|---|
| | Acc/MAE | Epoch/Total | Acc/MAE | Epoch/Total | Acc/MAE | Epoch/Total |
| GCN | 0.436±0.008 | 16.241 | 0.471±0.002 | 16.11hr | 0.443±0.003 | 15.83hr |
| GAT | 0.456±0.007 | 16.270 | 0.494±0.001 | 16.14hr | 0.463±0.009 | 16.04hr |
| GraphSage | **0.412±0.008** | 16.08hr | 0.444±0.005 | 16.05hr | 0.424±0.006 | 15.82hr |

## 5 EXPERIMENT RESULTS

The training is performed on a 2× Intel Xeon Silver 4210R, 64GB ram, and RTX 3090 Linux workstation. The loss function in training is mean square error (MSE), the adam optimizer [18] is used, the learning rate = 0.001, the Batch size is 32, and a total of 80 epochs are trained. The experiments are implemented in python 3.9 using the deep learning framework PyTorch [19] and the graph neural network framework PyTorch-geometric [20]. The data used for training and testing is taken from 22 circuits in the dataset OpenABC-D [7]. The characteristics of these 22 circuits are summarized in Table 4. Each circuit is run with K = 1500 optimization sequences, each of length L = 20, consisting of seven structural transformations of refactor, refactor -z, rewrite, rewrite -z, resub, resub -z, balance. The total size of the dataset is 33000, and the labels are the number of nodes after the circuit has been optimized.

A model that adopts a joint learning policy is trained with training data processed from a random selection of 80% of the optimization sequences for all circuits, and the model is tested for its ability to predict QoR given an unseen circuit-optimization sequence pair. This simulates the development of optimization sequences by experts for certain circuits, and the user needs to know how these optimization sequences will perform on other circuits, but without the time-consuming actual runs. Of this, 20% of the training data is used for validation.

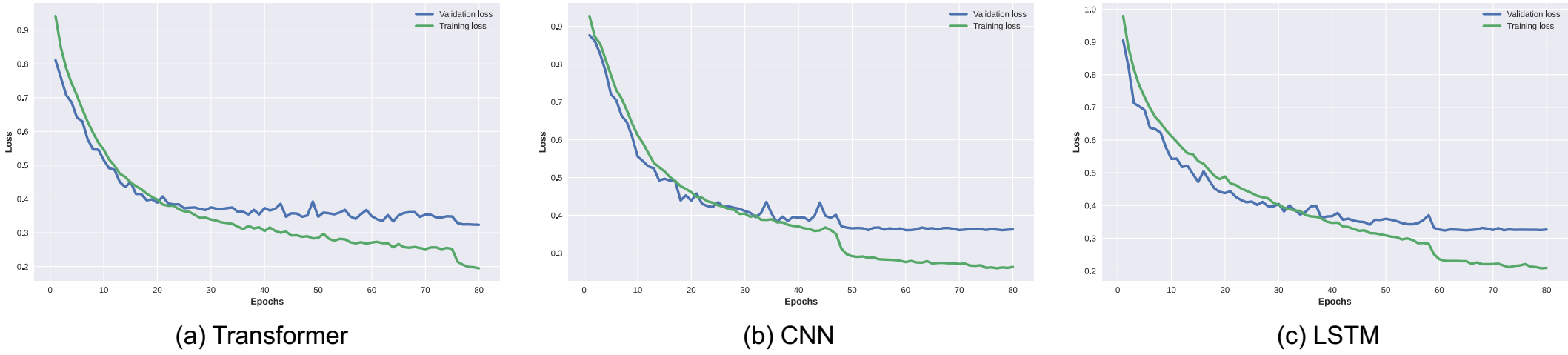

Figure 12: Training and validation curves of the joint policy based on the GraphSage.

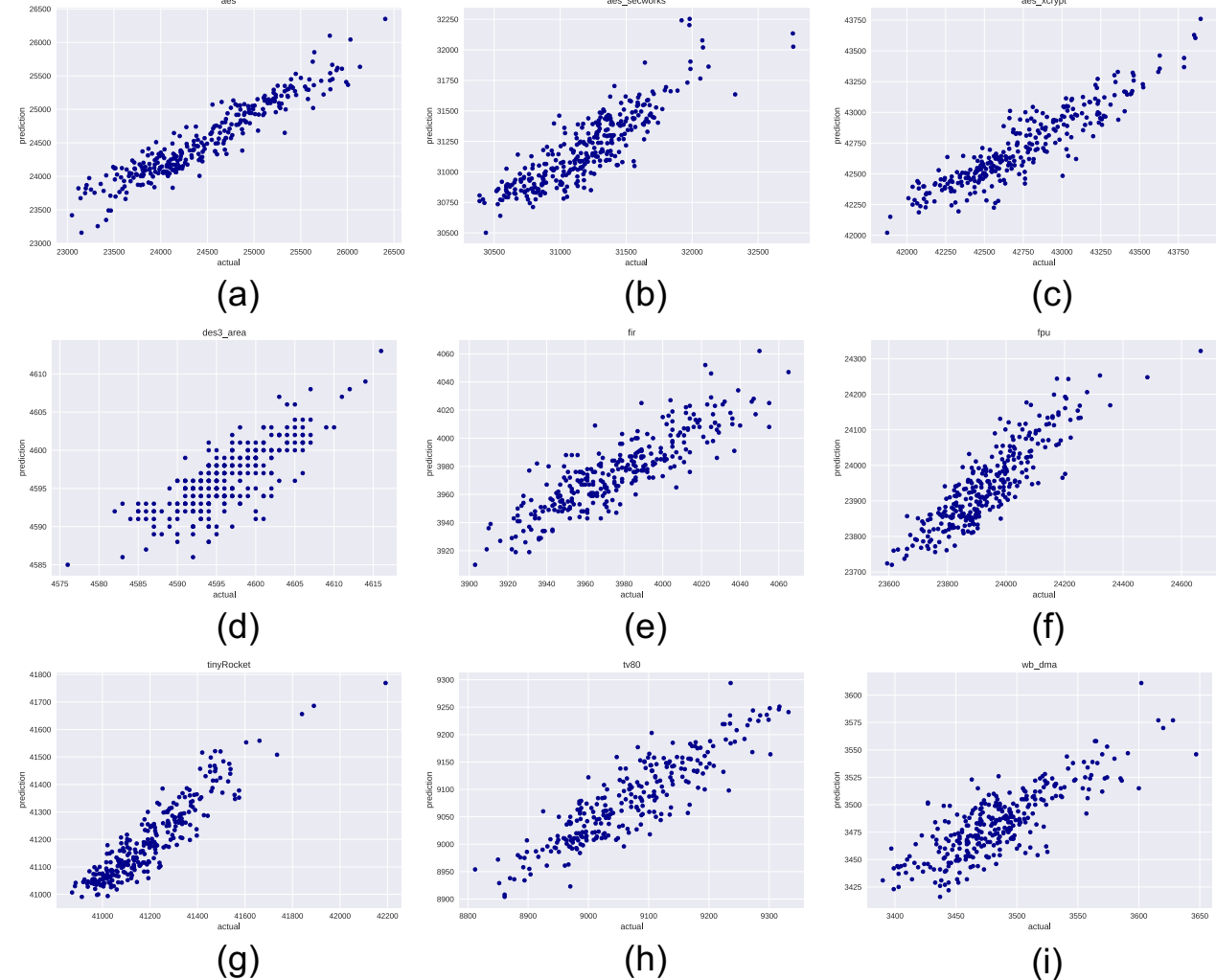

Figure 13: Visualization of part of the test results (unseen circuit-optimization sequence pairs).

Table 5 shows the results of nine combinations of optimization sequence feature and circuit feature extractor after adopting a joint learning policy. The combination of transformer and GraphSage achieves the best result with a mean absolute error (MAE) of 0.412. This demonstrates the usefulness of the self-attention mechanism for capturing optimization sequence features and the positive impact of learning aggregation functions for extracting circuit features. In contrast to the usual graph task, GAT obtained the worst score. We conjecture that the act of assigning different attention scores to each neighbor node does not learn the features of AIG well, and that the importance of its neighbors is perhaps determined by the input as buffer or inverter. This phenomenon will be thoroughly investigated in future work. In further, we note that the LSTM-based sequence feature extractor achieves slightly worse results than the transformer but with a shorter training time. This is since LSTM is lighter compared to the transformer, while the currently used circuits still lack large-scale industrial data and the coverage of circuit features is not comprehensive enough. Hence, using LSTM as a sequence feature extractor is a lighter choice in small-scale usage scenarios. However, the upper limit of the transformer is high, and the potential of the transformer will be exploited in future work using more comprehensive data. Figure 12 illustrates the training curve. Partial test results are visualized in Figure 13. The performance of the model varies for different circuits.

For a circuit like **aes** (Figure 13-a), the inference results are close to the true QoR values. But for a circuit like **des3_area** (Figure 13-d), the model cannot distinguish well between the QoR values of the optimization sequence, indicating that the data distribution during training is much different from the test distribution, and the model cannot make a good generalization for this type of circuit. Hence, a comprehensive collection of circuits with different feature distributions will be a priority in future work.

## 6 CONCLUSION

This work proposes a joint learning policy based on GNN and Transformer, that estimates the QoR of unseen circuit-optimization sequence pairs. This allows users to quickly know the optimization results of an optimization sequence for a circuit rather than performing time-consuming actual runs, reducing the difficulty of developing high-quality optimization sequences in a short time. To enable Transformer to parse optimization sequences, embedding methods for optimization sequences are proposed so that any optimization sequence can be represented as a vector with learning capabilities. In the proposed joint learning policy, three NLP feature extractors and GNNs are used respectively. The combination of Transformer and GraphSage achieves optimal performance with an MAE of 0.412. In addition, the combination of LSTM and graph neural network provides suboptimal results and can be a lighter alternative when the data size is not too large. Future work includes: 1. making a more comprehensive dataset of circuit features and exploiting the full potential of the transformer. 2. using GNNs that can learn edge features to fully extract the features of the circuit.

## ACKNOWLEDGMENTS

This work is supported by the National Natural Science Foundation of China under Grant 62131010 and Grant 61971389.

## REFERENCES

[1] Testa E, Soeken M, Amar L G, et al. Logic synthesis for established and emerging computing[J]. Proceedings of the IEEE, 2018, 107(1): 165-184.

[2] Brayton R, Mishchenko A. ABC: An academic industrial-strength verification tool[C]//International Conference on Computer Aided Verification. Springer, Berlin, Heidelberg, 2010: 24-40.

[3] Kahng A B. New directions for learning-based IC design tools and methodologies[C]//2018 23rd Asia and South Pacific Design Automation Conference (ASP-DAC). IEEE, 2018: 405-410.

[4] Haaswijk W, Collins E, Seguin B, et al. Deep learning for logic optimization algorithms[C]//2018 IEEE International Symposium on Circuits and Systems (ISCAS). IEEE, 2018: 1-4.

[5] Zhu K, Liu M, Chen H, et al. Exploring Logic Optimizations with Reinforcement Learning and Graph Convolutional Network[C]//2020 ACM/IEEE 2nd Workshop on Machine Learning for CAD (MLCAD). IEEE, 2020: 145-150.

[6] Hosny A, Hashemi S, Shalan M, et al. Drills: Deep reinforcement learning for logic synthesis[C]//2020 25th Asia and South Pacific Design Automation Conference (ASP-DAC). IEEE, 2020: 581-586.

[7] Chowdhury A B, Tan B, Karri R, et al. OpenABC-D: A Large-Scale Dataset For Machine Learning Guided Integrated Circuit Synthesis[J]. arXiv preprint arXiv:2110.11292, 2021.

[8] Yu C, Xiao H, De Micheli G. Developing synthesis flows without human knowledge[C]//Proceedings of the 55th Annual Design Automation Conference. 2018: 1-6.

[9] Yu C, Zhou W. Decision Making in Synthesis cross Technologies using LSTMs and Transfer Learning[C]//Proceedings of the 2020 ACM/IEEE Workshop on Machine Learning for CAD. 2020: 55-60.

[10] LeCun Y, Boser B, Denker J S, et al. Backpropagation applied to handwritten zip code recognition[J]. Neural computation, 1989, 1(4): 541-551.

[11] Hochreiter S, Schmidhuber J. Long short-term memory[J]. Neural computation, 1997, 9(8): 1735-1780.

[12] Vaswani A, Shazeer N, Parmar N, et al. Attention is all you need[C]//Advances in neural information processing systems. 2017: 5998-6008.

[13] Kipf T N, Welling M. Semi-supervised classification with graph convolutional networks[J]. arXiv preprint arXiv:1609.02907, 2016.

[14] Veličković P, Cucurull G, Casanova A, et al. Graph attention networks[J]. arXiv preprint arXiv:1710.10903, 2017.

[15] Hamilton W L, Ying R, Leskovec J. Inductive representation learning on large graphs[C]//Proceedings of the 31st International Conference on Neural Information Processing Systems. 2017: 1025-1035.

[16] Devlin J, Chang M W, Lee K, et al. Bert: Pre-training of deep bidirectional transformers for language understanding[J]. arXiv preprint

arXiv:1810.04805, 2018.
[17] Ba J L, Kiros J R, Hinton G E. Layer normalization[J]. arXiv preprint arXiv:1607.06450, 2016.
[18] Kingma D P, Ba J. Adam: A method for stochastic optimization[J]. arXiv preprint arXiv:1412.6980, 2014.
[19] Paszke A, Gross S, Massa F, et al. Pytorch: An imperative style, high-performance deep learning library[J]. Advances in neural information processing systems, 2019, 32: 8026-8037.
[20] Fey M, Lenssen J E. Fast graph representation learning with PyTorch Geometric[J]. arXiv preprint arXiv:1903.02428, 2019.
[21] Opencores hardware RTL designs. (https://opencores.org/).
[22] MIT Common Evaluation Platform(CEP). (https://github.com/mit-ll/CEP).
[23] Xie jian Jiang. AES 128/256-bit symmetric block cipher. (https://github.com/crypt-xie/XCryptCore/tree/master/ciphers/aes).
[24] Joachim Strömbergson and Olof Kindgren. AES 128/256-bit symmetric block cipher. (https://github.com/secworks/aes).
[25] Balkind J, McKeown M, Fu Y, et al. OpenPiton: An open source manycore research framework[J]. ACM SIGPLAN Notices, 2016, 51(4): 217-232.
[26] Ajayi T, Blaauw D. OpenROAD: Toward a self-driving, open-source digital layout implementation tool chain[C]//Proceedings of Government Microcircuit Applications and Critical Technology Conference. 2019.

## Appendix

本项目提出一种基于 Transformer 的优化序列与电路路径特征提取方法，通过联合学习优化序列和电路路径特征，实现对不同电路的优化序列 QoR 预测，提高逻辑综合工具的智能化水平和优化效果。

### （二）技术方案

本项目的方法主要包括以下模块：

### 1、数据集构建

将逻辑综合中的结构转换如：refactor， refactor -z, rewirte, rewrite -z, resub, resub -z, balance等组合排列成为优化序列，最大长度为$\boldsymbol{L_{Seq}}$，对于长度不足$\boldsymbol{L_{Seq}}$的优化序列使用零向量填充。

根据真实基准电路采样逻辑路径作为电路路径，使用零向量对长度小于最长路径的逻辑路径进行填充处理。

以每一个优化序列和电路路径作为一组配对的形式构建数据集，并同时设定标签；所述标签为，每个电路使用优化序列优化后的QoR(面积和延迟)。

### 2、优化序列转换

**(1) 优化命令的离散化**

- 定义优化操作词典，包括 7 个基本结构转换：

$O$ ={refactor, refactor -z, rewrite, rewrite -z, resub, resub -z, balance}

每个优化操作被赋予一个唯一索引：

**refactor→1, rewrite→3 , resub→5, …**

**（2）优化命令的嵌入**

将优化序列的索引通过所训练Transformer的嵌入层转换为向量：

如输入序列: [1, 2, 0, 6]

转换后: $e_1$，$e_2$, $e_0$, $e_6$] （每个$e_i$是一个n维向量）

**(3) 位置编码**

由于优化序列是有顺序的，本项目为优化序列中每个结构转换位置$i$计算**位置编码**，并与嵌入相加：

$$x_i = e_i + PE(i)$$

其中$PE(i)$ 通过如下预设公式计算，确保 Transformer 能感知命令顺序。

$$PE(i, 2j) = \sin\left(i/10000^{2j/d}\right), PE(i, 2j+1) = \cos\left(i/10000^{2j/d}\right)$$

**3、电路路径转换**

电路路径是从输入端到输出端的一条条逻辑路径，可以看作由逻辑单元组成的序列。例如：

[AND, AND, NOT, OR, AND]

可以按以下方式转换成 Transformer 能处理的格式。

**（1）逻辑门的离散化**

- 定义逻辑单元词典：

G={AND,OR,NOT,XOR,BUF}

- 每种逻辑单元映射为整数：

AND → 0, OR → 1, NOT → 2, XOR → 3, BUF → 4

**（2）电路路径的嵌入**

将电路路径经过所训练的Transformer的嵌入层转换为向量。

例如输入路径: [AND, AND, NOT, OR, AND]

转换后: $[v_0, v_0, v_2, v_1, v_0]$，其中$v_i$为一个n维向量。

**(3) 电路拓扑信息编码**

除了逻辑门类型，还需要考虑拓扑信息：

位置索引：记录该逻辑门在路径中的顺序,为每条逻辑路径中第i 个单元计算位置编码，并与嵌入相加：

$$x_i = v_i + PE(i)$$

其中$PE(i)$ 通过如下预设公式计算，确保 Transformer 能感知逻辑单元顺序。

$$PE(i, 2j) = \sin\left(i/10000^{2j/d}\right), PE(i, 2j+1) = \cos\left(i/10000^{2j/d}\right)$$

- Fan-in / Fan-out 编码：如果一个逻辑单元有多个输入或输出，它的拓扑特征也会影响 QoR。因此，对 Fan-in 和 Fan-out 作为输入送入一个共享 MLP，输出 d 维向量作为拓扑补偿项：

$$\mathrm{Topo}_i = \mathrm{MLP}([\ \text{Fan-in}_i,\ \text{Fan-out}_i])$$

最终的电路路径嵌入为：

$$x_i' = x_i + \mathrm{Topo}_i$$

**(4) 电路路径采样策略**

- **总采样数** $N_{\text{sample}}$ 设定为：

$$N_{\text{sample}} = min\left(T_{\text{total}}, max(500, 0.2 \times T_{\text{total}})\right)$$

  其中 $T_{\text{total}}$ 为电路的所有可能路径总数，小电路模块至少采样 500 条路径。大电路模块最多采样 20% 的路径，总量不超过上限。

- **路径采样比例**：
  - **短路径（L < 0.3 × $L_{nax}$）占 20%**，用于学习局部结构。
  - **典型路径（0.5 × $L_{med}$ ≤ L ≤ 1.0 × $L_{med}$）占 50%**，用于学习主要优化序列影响。
  - **长路径（L > 0.9 × L_max）占 30%**，确保 Transformer 关注关键长路径。

其中，$L_{nax}$代表路径的最大值，$L_{med}$代表路径的中位数。

- **不同长度路径适配方案：**

1）设定固定窗口大小$W_{\text{seg}}$(1024)：

1. • **长路径**（$L_k$ ＞ $W_{\text{seg}}$）→采用**滑动窗口分割**，分成多个 1024 长度的子路径。

2. • **短路径**（$L_k < W_{\text{seg}}$）→ 使用 **填充（Padding）** 填充到 1024，并在注意力计算时屏蔽填充部分。

2）路径分割与滑动窗口机制

3. 为了解决电路路径长度在不同电路间差异较大的问题，本项目在路径特征提取阶段引入了滑动窗口分割机制，以适配固定输入长度的 Transformer 模型。

4. 具体而言，对于路径长度$L_k$超过模型最大窗口大小$W_{\text{seg}}$的情况，本项目将该路径分割为多个连续、部分重叠的子路径序列，每个子路径长度为 $W_{\text{seg}}$。

5. 采用如下策略进行分割：

6. 设窗口长度为 $W_{\text{seg}}$，表示每次输入Transformer的最大路径长度；步长为 $S_{\text{stride}}$，表示滑动窗口在原始路径上的移动距离。对于路径总长度为 $L_k$ 的一条路径，分割过程如下：

- 第一个子路径从路径的第一个节点（索引 0）开始，提取连续$W_{\text{seg}}$个节点；
- 后续每一个子路径，其起始位置向后滑动固定步长$S_{\text{stride}}$，即从上一个子路径的起始位置加上$S_{\text{stride}}$得到；
- 具体地，第$i$个子路径的起始节点位置为：

7. $p_i = i \cdot S_{\text{stride}}$（其中 $i = 0,1,2,\ldots$）

8. 在每一步，均从当前位置$p_i$ 提取一个长度为 $W_{\text{seg}}$的片段，作为子路径输入模型。

9. 分割过程**持续进行，直到出现以下情形之一为止：**

1. 若当前位置$p_i$ 已经超过原始路径的末尾（即 $p_i \geq= L_k$），则终止；
2. 若从当前位置$p_i$ 向后提取$W_{\text{seg}}$个节点将超出路径末尾（即 $p_i + W_{\text{seg}} > L_k$），则仍然提取该子路径，不足部分以 零向量 进行填充；
3. 在完成该填充子路径之后，分割过程正式结束，不再继续滑动。

10.　该策略确保所有原始路径片段都被有效覆盖，且重叠区域可增强模型对前后门级结构的感知能力。推荐默认设置为步长 $S_{\text{stride}} = W_{\text{seg}}/2$，即每两个相邻子路径之间存在一半窗口长度的重叠，能够在精度与计算效率之间取得平衡。

**4、优化序列与电路特征的联合学习**

将2中所述的优化序列嵌入和3中所述的电路路径嵌入分别输入两个独立的Transformer。输出分别为为$F_{\text{seq}}$（优化序列特征）和$F_{\text{path}}$（电路路径结构特征）。将两者拼接为联合表示 $F_{\text{final}}$，供下游MLP预测模块使用。最终，利用 $F_{\text{final}}$输入由Transformer构成的回归模块，预测 QoR 指标（面积、延迟）。

本项目提供了一种基于Transformer的逻辑综合QoR双通道预测架构，旨在根据优化操作序列与电路路径结构的综合特征，准确预测逻辑综合后的关键 QoR（Quality of Result）指标，包括但不限于延迟与面积。与传统 GNN 方法相比，本项目基于 Transformer 的结构具备如下优势：能捕捉远距离操作之间的全局依赖（优化命令间，逻辑门间）。对于大电路具备更好的并行性与计算效率。使用统一的序列建模方式处理优化序列和路径结构，支持端到端训练和部署。

以下结合附图实施例对本项目作进一步详细描述。

本项目的方法流程图如图1所述，主要包括以下模块：

**S1：数据集构建**

将开源逻辑综合工具(ABC)中的结构转换如：refactor，　refactor -z, rewirte, rewrite -z, resub, resub -z, balance等组合排列成为优化序列，最大长度为$\boldsymbol{L_{Seq}}$，对于长度不足$\boldsymbol{L_{Seq}}$的优化序列使用零向量填充。

根据真实基准电路采样逻辑路径作为电路路径，使用零向量对长度小于最长路径的逻辑路径进行填充处理。

以每一个优化序列和电路路径作为一组配对的形式构建数据集，在逻辑综合工具ABC中对每个电路执行不同的优化命令组合，每组优化命令执行后，提取该电路的综合结果（如面积、时序等）作为标签。

对生成的数据集进行数据清洗和归一化处理，将处理好的数据集划分为70%训练集、10%验证集和20%测试集；

所述数据清洗包括与面积和延迟相关的异常值处理，将所有异常值都剔除数据集。其实所述异常值包括明显超出正常范围的值以及无效的值。

所述归一化为对标签进行Min-Max归一化，Min-max归一化的计算公式为:

$$x_{\text{norm}} = \frac{x - x_{min}}{x_{max} - x_{min}}$$

其中，$x$是原始数据值，$x_{max}$是每个电路中数据的最大值，$x_{min}$为每个电路中数据的最小值，$x_{\text{norm}}$为归一化后的数据值。需要额外说明的是，每个电路的$x_{min}$、$x_{max}$都是独立的，在计算$x_{\text{norm}}$时，使用其原始采样电路执行优化序列后得到的$x_{min}$和$x_{max}$。

**S2: 优化序列转换**

图2展示了逻辑综合结构转换变换为Transformer网络可处理格式的示意图，具体流程如下:

**(1) 优化命令的离散化**

- 定义优化操作词典，包括 7 个基本结构转换：

O={refactor, refactor -z, rewrite, rewrite -z, resub, resub -z, balance}

每个优化操作被赋予一个唯一索引：

**refactor→1, rewrite→3, resub→5, …**

**（2）优化命令的嵌入**

将优化序列的索引通过所训练Transformer的嵌入层转换为向量：

如输入序列: [1, 2, 0, 6]

转换后: $e_1$，$e_2$, $e_0$, $e_6$]（每个$e_i$是一个n维向量）

**(3) 位置编码**

由于优化序列是有顺序的，本项目为优化序列中每个结构转换位置 i 计算**位置编码**，并与嵌入相加：

$$x_i = e_i + PE(i)$$

其中$PE(i)$ 通过如下预设公式计算，确保 Transformer 能感知命令顺序。

$$PE(i,2j) = \sin\left(i/10000^{2j/d}\right), PE(i,2j+1) = \cos\left(i/10000^{2j/d}\right)$$

**S3: 电路路径转换**

电路路径是从输入端到输出端的一条条逻辑路径，可以看作由逻辑单元组成的序列。例如：

[AND, AND, NOT, OR, AND]

图3展示了电路路径转换为Transformer可处理格式的示意图，具体流程如下：

**（1）逻辑门的离散化**

- 定义逻辑单元词典：

G={AND,OR,NOT,XOR,BUF}

- 每种逻辑单元映射为整数：

AND → 0, OR → 1, NOT → 2, XOR → 3, BUF → 4

**（2）电路路径的嵌入**

将电路路径经过所训练的Transformer的嵌入层转换为向量。

例如输入路径: [AND, AND, NOT, OR, AND]

转换后: $[v_0, v_0, v_2, v_1, v_0]$，其中$v_i$为一个n维向量。

**(3) 电路拓扑信息编码**

除了逻辑门类型，还需要考虑拓扑信息：

位置索引：记录该逻辑门在路径中的顺序,为每条逻辑路径中第i 个单元计算位置编码，并与嵌入相加：

$$x_i = v_i + PE(i)$$

其中$PE(i)$ 通过如下预设公式计算，确保 Transformer 能感知逻辑单元顺序。

$$PE(i,2j) = \sin\left(i/10000^{2j/d}\right), PE(i,2j+1) = \cos\left(i/10000^{2j/d}\right)$$

- Fan-in / Fan-out 编码：如果一个逻辑单元有多个输入或输出，它的拓扑特征也会影响 QoR。因此，对 Fan-in 和 Fan-out 作为输入送入一个共享 MLP，输出 d 维向

量作为拓扑补偿项：

$$\mathrm{Topo}_i = \mathrm{MLP}([\,\text{Fan-in}\ _i,\ \text{Fan-out}\ _i])$$

最终的电路路径嵌入为：

$$x_i' = x_i + \mathrm{Topo}_i$$

**(4) 电路路径采样策略**

- **总采样数** $N_{\text{sample}}$ 设定为：

$$N_{\text{sample}} = min\left(T_{\text{total}}, max(500,0.2 \times T_{\text{total}})\right)$$

  其中 $T_{\text{total}}$ 为电路的所有可能路径总数，小电路模块至少采样 500 条路径。大电路模块最多采样 20% 的路径，总量不超过上限。

- **路径采样比例**：
  - **短路径（$L < 0.3 \times L_{nax}$）占 20%**，用于学习局部结构。
  - **典型路径（$0.5 \times L_{med} \le L \le 1.0 \times L_{med}$）占 50%**，用于学习主要优化序列影响。
  - **长路径（L > 0.9 × L_max）占 30%**，确保 Transformer 关注关键长路径。

其中，$L_{nax}$代表路径的最大值，$L_{med}$代表路径的中位数。

- **不同长度路径适配方案：**

3）设定固定窗口大小$W_{\text{seg}}$(1024)：

11. • **长路径**（$L_k > W_{\text{seg}}$）→ 采用**滑动窗口分割**，分成多个 1024 长度的子路径。

12. • **短路径**（$L_k < W_{\text{seg}}$）→ 使用 **零向量** 填充到 1024，并在注意力计算时屏蔽填充部分。

4）路径分割与滑动窗口机制

13. 为了解决电路路径长度在不同电路间差异较大的问题，本项目在路径特征提取阶段引入了滑动窗口分割机制，以适配固定输入长度的 Transformer 模型。

14. 具体而言，对于路径长度$L_k$超过模型最大窗口大小$W_{\text{seg}}$的情况，本项目将该路径分割为多个连续、部分重叠的子路径序列，每个子路径长度为 $W_{\text{seg}}$。

15. 采用如下策略进行分割：

16. 设窗口长度为 $W_{\text{seg}}$，表示每次输入Transformer的最大路径长度；步长为 $S_{\text{stride}}$ ，表示滑动窗口在原始路径上的移动距离。对于路径总长度为 $L_k$ 的一条路径，分割过程如下：

- 第一个子路径从路径的第一个节点（索引 0）开始，提取连续$W_{\text{seg}}$个节点；
- 后续每一个子路径，其起始位置向后滑动固定步长$S_{\text{stride}}$ ，即从上一个子路径的起始位置加上$S_{\text{stride}}$ 得到；
- 具体地，第$i$个子路径的起始节点位置为：

17. $p_i = i \cdot S_{\text{stride}}$ ( 其中 $i = 0,1,2, \dots$ )

18. 在每一步，均从当前位置$p_i$ 提取一个长度为 $W_{\text{seg}}$的片段，作为子路径输入模型。

19. 分割过程**持续进行，直到出现以下情形之一为止：**

4. 若当前位置$p_i$ 已经超过原始路径的末尾（即 $p_i \geq= L_k$），则终止；
5. 若从当前位置$p_i$ 向后提取$W_{\text{seg}}$个节点将超出路径末尾（即 $p_i + W_{\text{seg}} > L_k$ ），则仍然提取该子路径，不足部分以 零向量 进行填充；
6. 在完成该填充子路径之后，分割过程正式结束，不再继续滑动。

20. 该策略确保所有原始路径片段都被有效覆盖，且重叠区域可增强模型对前后门级结构的感知能力。推荐默认设置为步长 $S_{\text{stride}} = W_{\text{seg}}/2$，即每两个相邻子路径之间存在一半窗口长度的重叠，能够在精度与计算效率之间取得平衡。

**S4: 优化序列与电路特征的联合学习**

图4展示了联合学习优化序列与电路特征用于预测逻辑综合QoR的示意图。具体地，将优化序列$\mathcal{O} = [o_1, o_2, \ldots, o_T]$，经S2所述方法转换为优化序列嵌入向量。然后，将优化序列嵌入向量输入优化序列Transformer，提取优化操作语义特征，得到优化序列特征向量$F_{\text{seq}}$。将电路中采样得到的门级路径序列$\mathcal{P} = [p_1, p_2, \ldots, p_K]$经S3所述方法编码为嵌入向量。输入路径Transformer，提取电路结构特征，得到电路结构特征向量$F_{\text{path}}$。最终两个分支输出的高维特征向量$(F_{\text{seq}}, F_{\text{path}})$将拼接形成统一表示$F_{\text{final}}$，作为MLP预测模块的输入。模型最终输出电路延时预测值$\hat{y}_{\text{delay}}$和电路面积预测值$\hat{y}_{\text{area}}$。

将训练集输入所述基于Transformer的神经网络模型中，使用Adam优化器进行迭代训练，得到训练好的Transformer模型；具体是：

使用Adam优化器对整体损失函数$\mathcal{L}_{\text{reg}}$进行优化，整体的损失函数由延迟和面积联合回归，对每条训练样本 $i$，设真实值为$y_{\text{delay}}^{(i)}$、$y_{\text{area}}^{(i)}$，其预测值分别为$\hat{y}_{\text{delay}}^{(i)}$，$\hat{y}_{\text{area}}^{(i)}$，回归损失函数公式如下

$$\mathcal{L}_{\text{reg}} = \frac{1}{N}\sum_{i=1}^{N} \left[\lambda_d \cdot \left(y_{\text{delay}}^{(i)} - \hat{y}_{\text{delay}}^{(i)}\right)^2 + \lambda_a \cdot \left(y_{\text{area}}^{(i)} - \hat{y}_{\text{area}}^{(i)}\right)^2\right]$$

其中，$N$为输入的训练样本数量，$\lambda_d$，$\lambda_a$为损失加权系数, 默认$\lambda_d$=$\lambda_a$=1.0。

根据MLP预测模块输出的延迟值和面积值，即可获得电路在执行当前优化序列后的QoR结果。

本项目提供了一种基于Transformer的逻辑综合QoR双通道预测架构，旨在根据优化操作序列与电路路径结构的综合特征，准确预测逻辑综合后的关键 QoR（Quality of Result）指标，包括但不限于延迟与面积。本项目基于 Transformer 的结构具备如下优势：能捕捉远距离操作之间的全局依赖（优化命令间，逻辑门间）。对于大电路具备更好的并行性与计算效率。使用统一的序列建模方式处理优化序列和路径结构，支持端到端训练和部署。

**1．**一种基于 Transformer 架构用于预测逻辑综合结果质量（QoR）的方法，其特征在于，包括如下步骤：

步骤 S1: 结构数据集；具体是:

将逻辑综合结构转换排列组合为优化序列并应用于真实电路获得综合后的面积和延迟；

基于所述电路的门级连接图，对其进行路径采样，为每一个电路构造一组电路路径集，其中所述路径包括多个由逻辑门组成的节点序列；

对每条采样路径进行长度判断，若超出设定窗口上限，则采用滑动窗口机制对其进行分段处理，得到统一长度的子路径序列；

将每一个优化序列和电路路径集构建数据集，同时设定标签；所示标签包括该优化序列应用于该电路后的综合结果(面积和延迟)；

步骤 S2：对生成的数据集进行数据清洗和归一化处理，将处理好的数据集划分为训练集、验证集和测试集；

步骤 S3: 对优化序列进行词汇嵌入与位置编码，获得优化序列嵌入向量；

步骤 S4: 对每个逻辑门节点进行类型嵌入，并结合其 Fan-in、Fan-out 信息计算拓扑增强嵌入向量；

步骤 S5: 将所述拓扑增强嵌入与位置信息编码相加，形成电路路径的嵌入向量表示；

步骤 S6: 分别将所述优化操作序列与电路路径序列输入两个独立的 Transformer 网络，提取优化序列特征与电路结构特征；

步骤 S7: 将两个 Transformer 输出特征向量拼接，输入回归预测模块；

步骤 S8：预测该组合输入对应的逻辑综合结果指标，包括延迟与面积。

**2.** 根据权利要求 1 所示方法，其特征在于，步骤 S1 中所述路径采样包括：

对每个电路模块设其路径总数为 $T_{\text{total}}$，采样路径数 $N_{\text{sample}}$ 定义为：

$$N_{\text{sample}} = min\left(T_{\text{total}}, max(500, 0.2 \times T_{\text{total}})\right)$$

在满足总采样数约束的基础上，对路径按长度分布进行采样：

短路径（长度小于 $\mathbf{0.3 \times L_{nax}}$）占比 20%；

典型路径（长度落在 $\mathbf{0.5 \times L_{med}}$ 至 $\mathbf{1.0 \times L_{med}}$ 区间）占比 50%；

长路径（长度大于 $\mathbf{0.3 \times L_{nax}}$）占比 30%。

其中，$\boldsymbol{L_{nax}}$代表路径的最大值，$\boldsymbol{L_{med}}$代表路径的中位数。

**3.** 根据权利要求 1 所述的方法，其中每条采样路径若长度超过窗口上限 $W_{\text{seg}}$，则进行如下滑动窗口分段：

设置窗口长度 $W_{\text{seg}}$与步长 $S_{\text{stride}}$ ；

从路径起始位置开始，每隔 $S_{\text{stride}}$ 提取一个长度为 $W_{\text{seg}}$的子路径；

分段过程持续，直到下一段起始位置超过路径末尾为止；

最后一个子路径若不足 $W_{\text{seg}}$，则以零向量补足

**4.** 根据权利要求 3 所述的方法，其中步长设置为窗口长度的一半，即：

$$S_{\text{stride}} = W_{\text{seg}}/2$$

以确保相邻子路径存在一半长度的重叠区，用于增强上下文信息的连续性。

**5.** 根据权利要求 1 所述的方法，其中路径嵌入向量包括：

逻辑门类型嵌入向量$v_i$；

逻辑门位置信息编码$PE(i)$：

$$PE(i,2j) = \sin\left(i/10000^{2j/d}\right), PE(i,2j+1) = \cos\left(i/10000^{2j/d}\right)$$

拓扑结构的增强项：

$$\text{Topo}_i = \text{MLP}([\text{ Fan-in }_i, \text{ Fan-out }_i])$$

**6 .** 根据权利要求 1 所述的方法，其中预测模块为多任务回归结构，同时输出逻辑综合后的延迟与面积，其训练损失函数包括:

$$\mathcal{L} = \lambda_d \cdot \left(y_{\text{delay}} - \hat{y}_{\text{delay}}\right)^2 + \lambda_a \cdot (y_{\text{area}} - \hat{y}_{\text{area}})^2$$

其中，$y$与$\hat{y}$分别为真实值与预测值，$\lambda_d$、$\lambda_a$为可调加权系数。

**7.** 一种用于预测逻辑综合延迟与面积的智能预测系统，其特征在于，包括：

电路路径采样与预处理模块，用于根据路径长度分布和动态比例策略选取目标路径，并进行滑动窗口分段；

嵌入构造模块，用于构造逻辑门类型、拓扑结构与位置的综合嵌入向量；

双通道 Transformer 编码器，分别处理优化操作序列与电路路径序列；

回归预测模块，输出逻辑综合后的延迟与面积。